\documentclass[aps,pre,showpacs,preprint,nofootinbib]{revtex4-1}
\usepackage{chemarr}
\usepackage{amssymb}
\usepackage{graphicx}
\newcommand{\beq}{\begin{equation}}
\newcommand{\eeq}{\end{equation}}
\newcommand{\be}{\begin{eqnarray}}
\newcommand{\ee}{\end{eqnarray}}
\newcommand{\num}{1+\alpha \bar \nu}
\newcommand{\ab}{\alpha \beta}
\newcommand{\bb}{\bar \beta}
\newcommand{\bg}{\bar \gamma}
\newcommand{\bn}{\bar \nu}
\newcommand{\tg}{\zeta}
\begin{document}

%\preprint{}
%\begin{CJK*}{KS}{}
\title{Analytic stability analysis of three-component self-regulatory genetic circuit}
\author{Julian Lee}
%\author{Julian Lee (���ַ�)}
\email{jul@ssu.ac.kr}
\affiliation{Department of Bioinformatics and Life Science, Soongsil University, Seoul, Korea}
\date{\today}

\begin{abstract}
A self-regulatory genetic circuit, where a protein acts as a positive regulator of its own production, is known to be a simplest form of biological network with a positive feedback loop. Although at least three components, DNA, RNA, and the protein, are required to form such a circuit, the stability analysis of fixed points of the self-regulatory circuit has been performed only after reducing the system into a two-component system consisting of RNA and protein only, assuming a fast equilibration of the DNA component. Here, the stability of fixed points of the three-component positive feedback loop is analyzed by obtaining eigenvalues of full three dimensional Hessian matrix. In addition to rigorously identifying the stable fixed points and the saddle points, detailed information can be obtained, such as the number of positive eigenvalues near a saddle point. In particular, complex eigenvalues is shown to exist for sufficiently slow binding and unbinding of the auto-regulatory transcription factor to DNA, leading to oscillatory convergence to the steady states, a novel feature unseen in the two-dimensional analysis.
\end{abstract}
\pacs{87.10.Ed, 87.16.ad, 87.16.Yc, 87.18.Cf}

\maketitle
%\end{CJK*}

\section{Introduction}
Life is maintained by complex interaction of genes and proteins. The genes are turned on and off according to environmental change, so the gene network can be considered as a Turing machine with input and output. The genetic regulatory system is often modelled as a system of first-order differential equations that describes the time-dependent production and decay of various RNA and protein molecules, as well as binding of transcription factors to DNA that stimulate or inhibit the transcription of genes\cite{jong,smol,pigo,Radde,Angel,Keller,cir1,cir2,sn,pl,go,ci}.  The first step toward understanding such a system is to understand its steady state behavior. Depending on the network structure, the gene network exhibits multistability, meaning there can be more than one steady state of the protein concentrations\cite{Radde,Angel,Keller,sn,pl,go,ci,Haoge1}. Multistability not only forms the basis for explaining  the response of the cell to the environment, but can also be used to model epigenetics, where information other than genetic one are passed on to the next generation of a cell\cite{Keller}.  It has been shown that the existence of a positive feedback loop in the regulatory network is a necessary condition for a multistability\cite{sn,pl,go,ci}. Although general method for detecting multistability for a large class of systems with positive feedback loop has been proposed\cite{Angel}, it is of interest to perform an explicit analysis of a network if one can, where one can obtain more detailed information on the behavior of the system near the fixed point, such as the numbers of real and complex eigenvalues, and those with positive real parts in the case of a saddle point.

 The simplest biological system with a positive feedback loop is a self-regulatory genetic circuit where a protein acts as a transcription factor for its own gene, thus positively regulating its own production. According to the central dogma, there should be at least three components, DNA, RNA, and the protein. However, fixed-point analysis for such a simple self-regulatory genetic circuit has been performed only for two-dimensional system\cite{Stro,Griff,Gri68} or one-dimensional system\cite{Keller} which are obtained after dimensional reductions of the original set of equations, and it has been concluded that if the binding of the transcription factor to the DNA is cooperative, in the sense that more than two protein molecules are required to be bound in order to stimulate the production of their species, then there can be more than one possible steady-state concentrations of these proteins. 

Although fixed-points obtained from a dimensionally reduced system of equations are quite valid, the stability analysis of the fixed point in the dimensionally reduced system of equations is not justified: they are obtained under the assumption that some of the components, usually taken to be the DNA, are equilibrated to their steady-state values. The assumption of the fast-equilibration itself has to be tested, by showing that the eliminated variables are stable with respect to their steady-state values so that such a limit can indeed be taken. 

In this work, I consider the three-component auto-regulatory system consisting of DNA, RNA, and obtain analytic form for the eigenvalues of the Hessian at the fixed points, using the cubic root formula. Not only I show that this system can indeed exhibit bistability, but I can also explicitly investigate whether and how many of the eigenvalues have positive real values at the saddle point.  I also find that the eigenvalues at the stable fixed point can have imaginary parts, leading to oscillatory convergence of the concentrations to the steady state, a novel feature that cannot be seen in the two-dimensional analysis.

\section{The chemical equations}
The simplest system with a positive feedback loop is a self-regulatory genetic circuit where a protein $X$ binds to its gene to act as a transcription factor for its own species. Let us denote the inactive and active genes as $D$ and $D^*$, respectively. Also assume that $m$ molecules of proteins have to bind to the inactive gene in order to make it active and turn on the transcription. The active gene $D^*$ produces the RNA molecule $R$, which in turn produces molecules of the protein $X$ via translation mechanism. Both RNA and the protein degrade with time. This processes can be described by the following reactions: 
\be
D + mX \xrightleftharpoons[k_2]{k_1} D^*\nonumber\\
D^* \xrightarrow{a} D^* + R,\quad R \xrightarrow{b}  \varnothing \nonumber\\
R \xrightarrow{c}  X + R,\quad X \xrightarrow{d} \varnothing. \label{rateeq}
\ee
The rate equation for this process is 
\be
\lbrack \dot X \rbrack &=& -d [X] + c [R] + m k_2 [D^*] - m k_1 [D] [X]^m\nonumber\\
&=&  -d [X] + c [R]  + m (k_1 [X]^m + k_2) [D^*] - m k_1 T [X]^m \nonumber\\
\lbrack \dot R \rbrack &=&  -b [R] + a [D^*] \nonumber\\
\lbrack \dot D^* \rbrack &=& k_1 [D][X]^m - k_2 [D^*] \nonumber\\
&=& -(k_1 [X]^m + k_2) [D^*] + k_1 T [X]^m\label{deter}
\ee
where the condition that the concentration of DNA is a constant,
\be
[D] + [D^*] = T
\ee
was used to eliminate $[D]$ from the first and the third equation. Here, the concentration $[X]$ denotes the number density of the species $X$ divided by some fixed reference number density. Rigorously, it is the expectation value of the concentration. The distinction is often neglected when the fluctuation is small, but it is clear that the stochastic effect for DNA cannot be neglected, since typical gene copy number encoding a given protein is of order one per cell.  Assuming there is only one copy of the gene, $[D]$ and $[D^*]$, the expected concentrations of active and inactive genes, are also equal to the probabilities that a gene is in the off or on state, respectively, divided by the volume and the reference number density. As shown in the appendix A, the condition for the validity of Eq.(\ref{deter}) is 
\be
\langle N_D N_X^m \rangle = \langle N_D \rangle \langle N_X \rangle^m \label{decond}
\ee
where $N_D$ and $N_X$ denote the numbers of the protein $X$ and the active genes, respectively, and the brackets denote the expectation values. This condition will hold true if the fluctuation of the gene and that of the proteins are independent. In particular, it will hold if  the number of protein $X$ is large enough so that the relative fluctuation of its number is negligible. However, although this condition may hold at a certain point of time, the fluctuations of $N_X$ may grow in time and make the equation Eq.(\ref{deter}) invalid for later times. The full stochastic master equation or Fokker-Planck equation will have to analyzed to find the time-scale that the Eq.(\ref{decond}) is valid~\cite{Haoge1, Haoge2, Haoge3, Lip, Wangjin1,Wangjin2}, perhaps by numerical method, which is out of scope of the current manuscript. Here, I will simply assume the validity of Eq.(\ref{decond}) and analyze the consequence of the equation (\ref{deter}).  

Eq.(\ref{deter}) can be simplified by rescaling the time
\be
t \to \tau t,
\ee
and  defining the rescaled variables,
\be
x=[X]/\epsilon, y=[R]/\mu, z=[D^*]/T
\ee
and the parameters
\be
\alpha = b \tau, \beta = d \tau, \gamma =\tau k_2, \nu = m \gamma T/ \epsilon
\ee
where $\mu$, $\epsilon$, and $\tau$ are chosen to satisfy 
\be
k_1 \epsilon^m /k_2 = c \mu \tau \epsilon^{-1} = T \tau a \mu^{-1} =1,
\ee
to get the rescaled set of equations
\be
\dot x  &=& -\beta x +  y + \nu (z + z x^m - x^m) \nonumber\\
\dot y  &=&  -\alpha  y + z \nonumber\\
\dot z  &=&-\gamma (1 + x^m) z + \gamma x^m \label{3d}
\ee
which is the object of analysis in this work.

 Note that the in the first equation of Eq.(\ref{3d}), the first, second and the third terms describe the change of the protein number due to its degradation, production, and its binding and unbinding to the gene. Since only  $m$ molecules of the protein bind to the gene, in most cases of interest where the total number of proteins is much larger than $m$, the binding and unbinding process will have negligible contribution to the change of the total protein number. This can also be seen from the fact that $\nu$ is proportional to $T$, the concentration of gene which is usually much smaller than $[R]$ and $[X]$. Therefore $\nu$ is expected to be negligible in most practical applications. However, for the sake of generality, I will keep $\nu$ in the rest of the article. Also note that the $\nu$ dependent term becomes non-negligible near origin when only a few molecules of proteins are present, but in this limit the stochastic fluctuation of $[X]$ cannot be neglected and the deterministic equation (\ref{deter}) will no more be valid anyway(See also the appendix A).  
\section{The dimensional reduction and the two-dimensional system}
The fixed points of the system (\ref{3d}) is obtained by solving $\dot x = \dot y = \dot z$. One first solves $\dot z = 0$ to obtain 
\be
z=\frac{x^m}{1+x^m}.\label{zeq}
\ee
 Substituting into Eq.(\ref{3d}), we get
\be
\dot x  &=& -\beta x +  y  \nonumber\\
\dot y  &=&  -\alpha  y + \frac{x^m}{1 + x^m}   \label{2d}
\ee
The two-variable system described by Eq.(\ref{2d}) is the one investigated in the literature in detail\cite{Stro,Griff,Gri68}\footnote{Although a three-dimensional system has been considered in ref.\cite{Gri68}, whose fixed points coincide with those of Eq(\ref{2d}), it corresponds to the case where the first equation in Eq(\ref{2d}) is replaced by two linear equations, a rather unnatural situation where protein $x$ produces another protein that acts as a transcriptor for $x$.  The resulting equation is totally different from Eq(\ref{3d}) where only one species of protein is present and only the assumption of instantaneous equilibration of DNA is dropped.}.
 It is clear that although such a dimensional reduction is allowed for obtaining fixed points, one has to analyze the original system Eq.(\ref{3d}) in order to correctly analyze the stability of the fixed points. Eq(\ref{2d}) is based on the assumption that $\dot z = 0$ is always satisfied. This corresponds to the case when DNA equilibrates on a time scale much faster than that of the RNA or the protein. However, it has to be checked that the fixed point  is stable with respect to the perturbation orthogonal to the nullcline $\dot z = 0$, which is the main topic of interest in this work.
 
 In order to get the fixed points, we first solve the equation for the nullcline $\dot y = 0$, to get
\be
y=\frac{x^m}{\alpha(1+x^m)} \label{yeq}
\ee 
and substitute into the nullcline equation for $x$, $\dot x = 0$, which after some manipulation becomes
\be
x \left( \alpha \beta x^m -  x^{m-1} + \alpha \beta \right) = 0 \label{xeq}
\ee     
The values of $x$ for the fixed points are obtained by solving Eq.(\ref{xeq}), and those of $y$ and $z$ by Eq.(\ref{yeq}) and Eq.(\ref{zeq}). Since $x=0$ a solution of (\ref{xeq}), it is always a fixed point. Depending on the values of $\alpha$ and $\beta$, there can be a positive real fixed point $\tilde x$ that makes the second factor of Eq.(\ref{xeq}) zero:  
\be
\alpha \beta \tilde x^m -  \tilde x^{m-1} + \alpha \beta  = 0 \label{xeq2}
\ee    
The stability of the two-dimensional system is analyzed by linearizing the Eq.(\ref{2d}) around a fixed point $x_*$:
\be
\left(\begin{array}{c} \delta \dot x\\ \delta \dot y\end{array}\right) = {\bf A}_2 \left(\begin{array}{c} \delta x\\ \delta y\end{array}\right) = \begin{pmatrix} -\beta  & 1\\ \frac{m x_*^{m-1} }{(1+x_*^m)^2} & -\alpha \end{pmatrix} \left(\begin{array}{c} \delta x\\ \delta y\end{array}\right)\label{lin2}
\ee 
The details of the procedure are reviewed in appendix B. Here I just summarize the results:

\noindent
{\bf i) $m=1$}\\
$x=0$ is a unique stable fixed point when $\alpha \beta < 1$. For $\alpha \beta >1$, $x=0$ becomes an unstable fixed point, and there is an additional nonzero stable fixed point which is a root of Eq.(\ref{xeq2}). $\alpha \beta =1 $ is the marginal case where these two fixed points merge.

\noindent
{\bf ii) $m>1$}\\
$x=0$ is always a stable fixed point. There are additional nonzero fixed points $x_\pm$ with $x_- < x_+$ that correspond to the roots of Eq.(\ref{xeq2}) for 
\be
\left( \frac{m-1}{m \alpha \beta}\right)^m > m-1,\label{thecond}
\ee 
where $x_-$ and  $x_+$ are a saddle point and a stable fixed point, respectively. Therefore, $x=0$ is the unique stable fixed point for $\left(\frac{m-1}{m \alpha \beta}\right)^m < m-1$ whereas we get bistability for  $\left(\frac{m-1}{m \alpha \beta}\right)^m > m-1$. Again,  $\left(\frac{m-1}{m \alpha \beta}\right)^m = m-1$ corresponds to the marginal case where $x_+$ and $x_-$ merge into one.

In order to compare with those obtained from the three-dimensional analysis, we write the eigenvalues of the two-dimensional Hessian ${\bf A}_2$ in Eq.(\ref{lin2}): 
\be
\lambda_{\pm} &=&  \frac{1}{2}\left[ - \alpha - \beta \pm \sqrt{\left(\alpha - \beta \right)^2+ 4 \frac{m  x_*^{m-1}}{(1+x_*^m)^2} } \right] \label{ev}
\ee
where $x_*$ is the fixed point around which we linearized the equation. Note that since $x_* \ge 0$, the quantity inside the square root is non-negative, and therefore the eigenvalues are real. In case of stable fixed point where both of the eigenvalues are negative, this leads to asymptotically exponential convergence:
\be
x(t) &\to& x_* + u_{+x} e^{-\lambda_+ t} + u_{-x} e^{-\lambda_- t}\nonumber\\
y(t) &\to& y_* + u_{+y} e^{-\lambda_+ t} + u_{-y} e^{-\lambda_- t}
\ee
as $t \to \infty$, where ${\bf u}_\pm$ are the eigenvectors corresponding to $\lambda_\pm$, with $u_{\pm x}$ and $u_{\pm y}$ denoting their $x$ and $y$ components.  However, as will be shown later, the analysis of the full three-dimensional system Eq.(\ref{3d}) reveals that oscillatory convergence to a stable fixed point is also possible when the conversion between the active and inactive gene is slow enough.
\section{The stability analysis of the three-dimensional system}
As noted earlier, the stability analysis of the two-dimensional system given by Eq.(\ref{2d}) is rather incomplete. Considering a small deviation of $x$, $y$, and $z$ around a fixed point $(x_*,y_*,z_*)$ of Eq (\ref{3d}), with $x=x_* + \delta x$,$y=y_* + \delta y$, and $z=z_* + \delta z$, we get a linearized equation
\be
\left(\begin{array}{c} \delta \dot x\\ \delta \dot y\\ \delta \dot z\end{array}\right) = {\bf A}_3 \left(\begin{array}{c} \delta x\\ \delta y\\ \delta z\end{array}\right) = \begin{pmatrix} -\beta -\nu m x_*^{m-1} (1-z_*)  & 1 & \nu (1 + x_*^m)\\0 & -\alpha & 1\\\gamma m x_*^{m-1} (1-z_*) & 0 & -\gamma (1 + x_*^{m})\end{pmatrix} \left(\begin{array}{c} \delta x\\ \delta y\\ \delta z\end{array}\right)
\ee 
Solving the characteristic equation for the matrix ${\bf A}_3$,
\be
|\lambda - {\bf A}_3| &=& (\lambda + \alpha) (\lambda + \beta + \nu m x_*^{m-1}(1-z_*))(\lambda + \gamma  (1+x_*^m))\nonumber\\
&&- \gamma m x_*^{m-1}(1-z_*) -  (1+x_*^m) (\lambda+\alpha) \gamma\nu  m x_*^{m-1} (1-z_*)\nonumber\\
&=& \lambda^3 + (\alpha + \bar \beta + \bar \gamma) \lambda^2 + (\alpha \bar \beta + \bar \beta \bar \gamma + \bar \gamma \alpha -  \tg \bar \nu ) \lambda + (\alpha \bar \beta \bar \gamma - \tg (1 + \alpha \bar \nu)) = 0,\label{3ch}
\ee 
where
\be
\bb &\equiv& \beta + m x_*^{m-1} (1-z_*) \nu = \beta + \frac{\nu m x_*^{m-1}}{ x_*^m + 1}\nonumber\\
\bg &\equiv& \gamma (1+x_*^m) \nonumber\\
\tg &\equiv& \gamma m x_*^{m-1} (1 - z_*) = \gamma \frac{m x_*^{m-1}}{1 + x_*^m} \nonumber\\
\bn &\equiv& \nu (1 + x_*^m)
\ee
The case of $m>1$ and $x_* = y_* = z_* = 0$ is the easiest to analyze. In this case, $\bb=\beta$, $\bg=\gamma$, $\bn=\nu$, $\zeta=0$, and the solutions to the characteristic equation Eq.(\ref{3ch}) are easily shown to be
\be
\lambda = - \alpha, -\beta, - \gamma,
\ee
all of which are negative. Therefore, for $m>1$, the origin is always a stable fixed point, as was the case for the two-dimensional system. It remains to analyze the other cases.

In general, the three roots of a cubic equation
\be
\lambda^3 + p_2 \lambda^2 + p_1 \lambda + p_0 =0
\ee
are given by\cite{Poly}
\be
\lambda_0 &=& -\frac{p_2}{3}+\frac{1}{3}\sqrt[3]{\frac{R+\sqrt{R^2-4 Q^3}}{2}} + \frac{1}{3}\sqrt[3]{\frac{R-\sqrt{R^2-4 Q^3}}{2}}\nonumber\\
\lambda_\pm &=& -\frac{p_2}{3}+\frac{e^{\mp 2 \pi i/3}}{3}\sqrt[3]{\frac{R+\sqrt{R^2-4 Q^3}}{2}} + \frac{e^{\pm 2 \pi i/3}}{3}\sqrt[3]{\frac{R-\sqrt{R^2-4 Q^3}}{2}}\label{3root}
\ee
where 
\be
Q &\equiv& p_2^2-3 p_1 \nonumber\\
R &=& -2 p_2^3 + 9 p_1 p_2 - 27 p_0.
\ee
When the coefficients $p_0$,$p_1$, and $p_2$ are all real, we either get three real roots, or one real root and a conjugate pair of complex roots, depending on the sign of
\be
R^2-4 Q^3 = -27 \Delta
\ee 
where
\be
\Delta \equiv 18 p_0 p_1 p_2 - 4 p_2^3 p_0 + p_2^2 p_1^2 - 4 p_1^3 -27 p_0^2. 
\ee
When $\Delta > 0$, $\sqrt{-27 \Delta}=\sqrt{R^2-4 Q^3}$ is imaginary, and consequently the second and the third terms of each line of Eq.(\ref{3root}) form conjugate pair, so all three roots are real. On the other hand, if $\Delta \le 0$, $\lambda_0$ is real, but the second and the third terms of $\lambda_\pm$ are linear combinations of distinct real quantities with complex coefficients $e^{\pm 2 \pi i/3}$, so $\lambda_\pm$ form a conjugate pair of complex roots.

For the characteristic equation Eq.(\ref{3ch}) we get
\be
R &=& - 2 (\alpha + \bb + \bg)^3 + 9(\alpha + \bb + \bg)(\alpha \bb + \bb \bg + \bg \alpha - \zeta \bar \nu)\nonumber\\
&& - 27 \left(\alpha \bb \bg - \tg (\num)\right)\nonumber\\
&=& (\alpha + \bb - 2 \bg)(\bb + \bg - 2 \alpha)(\bg + \alpha - 2 \bb) + 27 \tg - 9 (\bb + \bg - 2 \alpha) \tg \bn \label{Del}
\ee
and
\be
\Delta &=& -\frac{R^2-4 Q^3}{27} = 18 (\alpha + \bb + \bg)(\alpha \bb + \bb \bg + \bg \alpha - \bn \tg)(\alpha \bar \beta \bar \gamma - \zeta (1 + \alpha \bar \nu) )\nonumber\\
&&-4(\alpha + \bb + \bg)^3 (\alpha \bb \bg - \tg (\num))\nonumber\\
&&+(\alpha + \bb + \bg)^2 (\alpha \bb + \bb \bg + \bg \alpha - \bn \tg)^2\nonumber\\
&&-4 (\alpha \bb + \bb \bg + \bg \alpha - \bn \tg)^3\nonumber\\
&&-27 (\alpha \bb \bg -(\num) \tg)^2\nonumber\\
&=& (\alpha - \bb)^2 (\bb-\bg)^2 (\bg-\alpha)^2 - 2 (\alpha + \bb - 2 \bg)(\bb+\bg-2\alpha)(\bg+\alpha-2\bb)\tg\nonumber\\
&& -27 \tg^2 + 18 (\bb+\bg- 2 \alpha) \tg^2 \bn\nonumber\\
&& + 2 (\alpha - \bb)(\alpha - \bg) (2 \alpha^2 - \bb^2 - \bg^2 - 2 \alpha \bb + 4 \bb \bg - 2 \bg \alpha) \tg \bn\nonumber\\
&& + (-8 \alpha^2 + \bb^2 + \bg^2 + 8 \alpha \bb + 8 \bg \alpha - 10 \bb \bg ) \tg^2 \bn^2 + 4 \tg^3 \bn^3 \label{R}
\ee 
which will be used for further analysis below.
\subsection{Limit of instantaneous equilibration of the gene}
The limit of instantaneous equilibration of the gene is obtained by taking the limit of $\gamma \to \infty$. The leading order expansions of  $R$  and $\Delta$ are
\be
R &=& -2 \bg^3 + 3 (\alpha + \bb) \bg^2 - 9  \tg \bn  \bg + O(\gamma)\nonumber\\
 &=& - 2 \gamma^3 (1 + x_*^m)^3 + \gamma^2 \left( 3 (1+x_*^m)^2(\alpha + \beta)- 6 m \nu (1 + x_*^m) x_*^{m-1} \right) + O(\gamma)\nonumber\\
\Delta &=& (\alpha-\bb)^2 \bar \gamma^4 + 4 \tg \bg^3 + 2 (\alpha-\bb)\bg^3 \tg \bn + \bn^2 \tg^2 \bg^2 + O(\gamma^3)  \nonumber\\
&=& \gamma^4 (1 + x_*^m)^2 \left( (1 + x_*^m)^2 (\alpha - \beta)^2 + 4 m x_*^{m-1}\right) + O(\gamma^3) .
\ee
We see that since the coefficient of $\gamma^4$ in the expansion for $\Delta$ is positive, $\Delta > 0$ and consequently all the eigenvalues are real when $\gamma$ is large enough.
In this case, we also have
\be
&&\frac{1}{2}(R \pm \sqrt{R^2 - 4 Q^3}) = \frac{1}{2}(R \pm \sqrt{-27 \Delta}) \nonumber\\
&=& -\gamma^3 (1 + x_*^m)^3 + \gamma^2 \left(\frac{3}{2}(1 + x_*^m)^2 (\alpha + \beta) - 3 m \nu x_*^{m-1} \right) \nonumber\\
&& \pm i \frac{ 3 \sqrt{3}}{2} \gamma^2 ( 1 + x_*^m) \sqrt{(\alpha-\beta)^2(1+x_*^m)^2 + 4 m x_*^{m-1}} + O(\gamma)
\ee
and consequently
\be
\left(\frac{1}{2}(R \pm \sqrt{-27 \Delta})\right)^{1/3}
&=& -\gamma (1 + x_*^m) \big[1 + \frac{1}{\gamma} \Big(-\frac{3}{2}\frac{\alpha + \beta}{1 + x_*^m} + 3 \frac{m \nu x_*^{m-1}}{(1 + x_*^m)^3}  \nonumber\\
&& \mp i \frac{3 \sqrt{3}}{2 (1 + x_*^m)^2} \sqrt{(\alpha-\beta)^2(1+x_*^m)^2 + 4 x_*^{m-1}}\Big) + O(\gamma^{-2})\big]^{1/3}\nonumber\\
&=& -\gamma (1 + x_*^m) \big[1 + \frac{1}{\gamma} \Big(-\frac{1}{2}\frac{\alpha + \beta}{1 + x_*^m} + \frac{m \nu x_*^{m-1}}{(1 + x_*^m)^3}  \nonumber\\
&& \mp i \frac{\sqrt{3}}{2 (1 + x_*^m)^2} \sqrt{(\alpha-\beta)^2(1+x_*^m)^2 + 4x_*^{m-1} }\Big) + O(\gamma^{-2})\big]\nonumber\\
&=& -\gamma (1 + x_*^m) + \frac{1}{2}(\alpha + \beta) - \frac{m \nu x_*^{m-1}}{(1 + x_*^m)^2}  \nonumber\\
&& \pm i \frac{\sqrt{3}}{2} \sqrt{(\alpha-\beta)^2 + \frac{4x_*^{m-1}}{(1+x_*^m)^2} } + O(\gamma^{-1}).
\ee
Therefore, from Eq.(\ref{3root}) we have
\be
\lambda_0 = -\frac{\alpha + \bb + \bg}{3} +\frac{2}{3}\operatorname{Re}\left(\left(\frac{1}{2}(R + \sqrt{-27 \Delta})\right)^{1/3}\right)=-\gamma (1 + x_*^m) + O(1)
\ee
which is real and negative, and goes to $-\infty$ as $\gamma \to \infty$. This shows that the trajectory of the system zaps onto the nullcline $\dot z = 0$ as $\gamma \to \infty$, justifying the two-dimensional system Eq.(\ref{2d}) in such a limit.
We also have
\be
&&\operatorname{Re}\left[e^{2 \pi i/3}\left(\frac{R \pm \sqrt{-27 \Delta}}{2}\right)^{1/3}\right]\nonumber\\
&=&\operatorname{Re}\Bigg[\left( \frac{-1+i \sqrt{3}}{2}  \right)\bigg(-\gamma (1 + x_*^m) + \frac{1}{2}(\alpha + \beta) - \frac{m \nu x_*^{m-1}}{(1 + x_*^m)^2}  \nonumber\\
&& \pm i \frac{\sqrt{3}}{2} \sqrt{(\alpha-\beta)^2 + \frac{4x_*^{m-1}}{(1+x_*^m)^2} } + O(\gamma^{-1})\bigg)\Bigg]\nonumber\\
%\left(-\gamma (1 + x_*^m)\right) \big[1 + \frac{1}{\gamma} \Big(-\frac{1}{2}\frac{\alpha + \beta}{1 + x_*^m} + \frac{m \nu x_*^{m-1}}{(1 + x_*^m)^3}  \nonumber\\
%&& \mp i \frac{\sqrt{3}}{2 (1 + x_*^m)^2} \sqrt{(\alpha-\beta)^2(1+x_*^m)^2 + 4x_*^{m-1} }\Big) + O(\gamma^{-2})\big] \nonumber\\
%&=& \frac{\gamma (1 + x_*^m)}{2} \big[1 + \frac{1}{\gamma} \Big(-\frac{1}{2}\frac{\alpha + \beta}{1 + x_*^m} + \frac{  m \nu x_*^{m-1}}{(1 + x_*^m)^2}\Big)  \nonumber\\
%&& \mp \frac{\sqrt{3}}{2 (1 + x_*^m)^2 \gamma} \sqrt{(\alpha-\beta)^2(1+x_*^m)^2 + 4 x_*^{m-1}} + O(\gamma^{-2})\big]\nonumber\\
&=& \frac{\gamma (1 + x_*^m)}{2}  -\frac{\alpha + \beta}{4} + \frac{  m \nu x_*^{m-1}}{2(1 + x_*^m)}  \nonumber\\
&& \mp \frac{3}{4  } \sqrt{(\alpha-\beta)^2 + 4 \frac{x_*^{m-1}}{(1+x_*^m)^2}} + O(\gamma^{-1}),
\ee
leading to
\be
\lambda_\pm &=& -\frac{\alpha + \bb + \bg}{3} +\frac{2}{3}\operatorname{Re}\left[\left( \frac{-1+i \sqrt{3}}{2}  \right)\left(\frac{1}{2}(R \mp \sqrt{-27 \Delta})\right)^{1/3}\right]\nonumber\\
&=&
\frac{1}{2} \left[ -\alpha - \beta \pm \sqrt{(\alpha - \beta)^2 + \frac{ 4 m x_*^{m-1}}{(1 + x_*^m)^2}} \right] + O(\gamma^{-1})
\ee
which agree with Eq.(\ref{ev}).
\subsection{General stability analysis of the fixed points}
A fixed point is stable if the real parts of three eigenvalues are all negative, whereas it is a saddle point if some of the them have positive real parts. By utilizing the root formula for a cubic equation, one can not only determine the stability of the fixed points, but also obtain more detailed information such as the imaginary parts and the signs of the eigenvalues.
 
Denoting the three eigenvalues as $\lambda_i (i = 1,2,3)$, we note that
\be
\sum_i \lambda_i &=& - \alpha - \beta -\gamma (1 + x_*^m) - \nu \frac{ m x_*^{m-1}}{1 + x_*^m} < 0 \nonumber\\ 
\sum_{i < j} \lambda_i \lambda_j &=& \alpha \beta + \beta \gamma (1 + x_*^m) + \gamma (1 + x_*^m) \alpha + \alpha \frac{\nu m x_*^{m-1}}{1 + x_*^m} > 0 \nonumber\\
\prod_i \lambda_i &=& \gamma \left( \frac{m x_*^{m-1}}{1+x_*^m} - \alpha \beta (1 + x_*^m) \right). \label{coeff}
\ee
The first line implies that at least one eigenvalue has a negative real part, which is also a {\it necessary} condition for a fixed point being stable. 
When $\prod_i \lambda_i > 0$, the realities and signs of the eigenvalues are either $(-,-,+)$ or $(C,C^*,+)$, where $+$ and $-$ denote positive and negative real roots respectively, and $C,C*$ denote a conjugate pair of complex roots. Also, since at least one eigenvalue must have negative real part due to the first equation of Eqs.(\ref{coeff}), the real parts of the complex conjugate roots are negative. 

On the other hand, if $\prod_i \lambda_i < 0$, the realities and signs are either $(-,-,-)$ or $(-,C,C^*)$. Note that $(-,+,+)$ is excluded due to the second line of Eq (\ref{coeff}). If $\lambda_3 < 0$ and $\lambda_1,\lambda_2 >0$, we have
\be
\lambda_1 \lambda_2 > |\lambda_3| (\lambda_1 + \lambda_2) > (\lambda_1 + \lambda_2)^2 \label{lineq}
\ee
where the last inequality follows from the first line of Eq.(\ref{coeff}), which is
\be
|\lambda_3| > \lambda_1 + \lambda_2.
\ee
Since 
\be
(\lambda_1 - \lambda_2)^2 = (\lambda_1 + \lambda_2)^2 - 4 \lambda_1 \lambda_2 \ge 0,
\ee
we get from Eq.(\ref{lineq})
\be
\lambda_1 \lambda_2 > 4 \lambda_1 \lambda_2
\ee
which is an obvious contradiction. Therefore, the signs of the eigenvalues cannot be $(-,+,+)$. The signs of the real parts of the complex roots remain to be determined. Note that two of the roots are complex if and only if $-27 \Delta = R^2 - 4 Q^3 > 0$. Also note that
\be
Q &=& \alpha^2 + \bb^2 + \bg^2 - \alpha \bb - \bb \bg - \bg \alpha + 3 \bn \tg \nonumber\\
&=& \frac{1}{2}\left[ (\alpha -\bb)^2 + (\bb-\bg)^2 + (\bg - \alpha)^2\right] + 3 \bn \tg > 0,
\ee
and consequently $\sqrt{R^2-4 Q^3} < R$, so that the sign of $[\frac{R \pm \sqrt{R^2 - 4 Q^3}}{2}]^{1/3}$ is the same as that of $R$. From Eqs.(\ref{Del}) and (\ref{R}), we see that
\be
\Delta &=& - 2 \tg R +27 \tg^2 + \left(4 \tg \bn + (\bb-\bg)^2\right)\left(\tg \bn - (\alpha-\bb)(\alpha-\bg)\right)^2. 
\ee
Since the coefficient of $R$ is negative and the remaining term is positive, we see that if $\Delta < 0$ so that there are complex roots, then $R>0$. Since the real part of the complex roots are
\be
\frac{1}{3}\left[ -(\alpha + \bb + \bg)-\frac{1}{2}\left(\frac{R+\sqrt{R^2-4Q^3}}{2}\right)^{1/3} -\frac{1}{2}\left(\frac{R-\sqrt{R^2-4Q^3}}{2}\right)^{1/3} \right]
\ee
which are sum of three negative numbers, the real parts of the complex roots are negative. 

In summary, we get a saddle point for $\prod_i \lambda_i > 0$  where the three roots of the Hessian are of the form $(-p,-q,r)$ or $(-p+iq,-p-iq, r)\quad (p,q,r > 0)$. On the other hand we get a stable fixed point for $\prod_i \lambda_i <0$ where the roots are of the form $(-p,-q,-r)$ or $(-p+iq,-p-iq,-r)\quad (p,q,r > 0)$. $\prod_i \lambda_i = 0$ corresponds to the marginal case where the roots are of the form $(-p,-q,0)$ or $(-p+iq,-p-iq,0)\quad (p,q > 0)$.

Now we analyze the stability of each fixed point by determining the sign of $\prod_i \lambda_i$.  For $m=1$ and $x_*=0$, we have 
\be
\prod_i \lambda_i = \gamma ( 1 - \alpha \beta)
\ee
Therefore, when $\alpha \beta >1$ so that the origin is the unique fixed point, $\prod_i \lambda_i < 0$, so it is a stable fixed point, whereas when $\alpha\beta < 1 $ so that there is an additional nonzero fixed point, $\prod_i \lambda_i > 0$  and the origin becomes a saddle point.

For additional nonzero fixed points, we have
\be
\prod_i \lambda_i &=& \gamma ( \frac{m x_*^{m-1}}{1 + x_*^m} - \alpha \beta ( 1 + x_*^m) )\nonumber\\
&=& \gamma ( \frac{m x_*^{m-1}}{1 + x_*^m} - x_*^{m-1} )\nonumber\\
&=& \frac{\gamma  x_*^{m-1}}{1 + x_*^m}(m-1-x_*^m)
\ee
where Eq.(\ref{xeq2}) was used when going from the first to the second line. For $m=1$, it is negative, so the additional nonzero fixed point is stable whenever it exists. For $m>1$, it remains to determine the sign of
\be
m-1 - x_*^m.
\ee 
Note that the function
\be
g(x) \equiv m-1-x^m
\ee
vanishes at
\be
x = x_0 \equiv (m-1)^{1/m}.
\ee
Substituting $x_0$ into 
\be
f(x) \equiv x^m - \frac{x^{m-1}}{\alpha\beta} + 1,
\ee
we get
\be
f(x_0) = m - \frac{1}{\alpha\beta} (m-1)^{(m-1)/m}.
\ee
Whenever the nonzero fixed points exist, the condition for which is Eq.(\ref{thecond}), $f(x_0)<0$. Since $f(0)>0$ and $\lim_{x \to \infty} f(x) > 0$, we see that $x_- < x_0 < x_+$. Since $g(x)$ is a monotonically decreasing function of $x$, we see that $g(x_-)>0$ and $g(x_+)<0$. Therefore, $x_-$ is a saddle point and $x_+$ is a stable fixed point.

Thus, we rigorously identified the stability of the fixed points in the three-dimensional system of equations Eq.(\ref{3d}). For $m=1$, when the degradation rates are large compared to production rates, $\alpha \beta > 1$, then $x_*=y_*=z_*=0$ is the unique stable fixed point. When $\alpha \beta < 1$, the origin becomes an unstable fixed point, and the unique stable fixed point appears at  nonzero values of the concentrations, at $(x_*,y_*,z_*)=(1/\alpha\beta-1, 1/\alpha-\beta, 1-\alpha\beta)$. Since there is a unique stable fixed point regardless of the parameters,  we have no bistability for $m=1$. On the other hand, for $m>1$, $x_*=y_*=z_*=0$ is a stable fixed point regardless of the parameter values. For  $((m-1)/m \alpha \beta)^m < m-1$, this is the unique stable fixed point. For $((m-1)/m \alpha \beta)^m > m-1$ we get an additional stable fixed point $x_+$ and a saddle point $x_-$ which are the roots to the equation (\ref{xeq2}), leading to bistability.
\subsection{Complex eigenvalues and the oscillatory behavior}
It is to be noted that, in contrast to the two-dimensional system, the eigenvalues of the full three dimensional systems are allowed to have complex values, in the region of the parameter space where $\Delta <0$.  To be specific, let us consider the Hessian around a stable fixed point. The eigenvector $u_0$ of ${\bf A}_3$ for real eigenvalue $\lambda_0 = - r \quad (r>0)$ is also a real vector, but when $\Delta<0$ so that $\lambda_\pm$ for a conjugate pair, the corresponding eigenvectors $u_\pm$ also form a conjugate pair. Denoting $\lambda_\pm = -p \pm iq\quad (p>0)$, we construct real vectors
\be
u_1&=& u_+ + u_-\nonumber\\
u_2&=& \frac{1}{i}(u_+ - u_-) 
\ee
so that
\be
{\bf A}_3  u_1 =  -p u_1 - q u_2,\quad {\bf A}_3  u_2 = -p u_2 + q u_1. 
\ee 
The parameter $p$ is the rate of approach to the fixed point along the plane spanned by $u_1$ and $u_2$, whereas $q$ is the rate of oscillation between the vectors $u_1$ and $u_2$. 

In the limit of $t \to \infty$, the behavior of the variables $x$, $y$, and $z$ are thus
\be
\left(\begin{array}{c} x(t) \\ y(t) \\ z(t)\end{array}\right) = \left(\begin{array}{c} x_* + A u_{0x} e^{-r t} + B e^{-p t} \left( u_{1x}  \cos (q t+\phi) - u_{2x}   \sin (q t+\phi)\right ) \\ y_* + A u_{0y} e^{-r t} + B e^{-p t} \left( u_{1y}  \cos (q t+\phi) - u_{2y}   \sin (q t+\phi)\right ) \\ z_* + A u_{0z} e^{-r t} + B e^{-p t} \left( u_{1z}  \cos (q t+\phi) - u_{2z}   \sin (q t+\phi)\right )\end{array}\right)
\ee
where $u_{ij}$ denotes the $j$-th component of the eigenvector ${\bf u}_i$, and the coefficients $A$,$B$, and the phase $\phi$ depend on the initial condition.
 
\begin{figure}
\includegraphics[width=5.0cm]{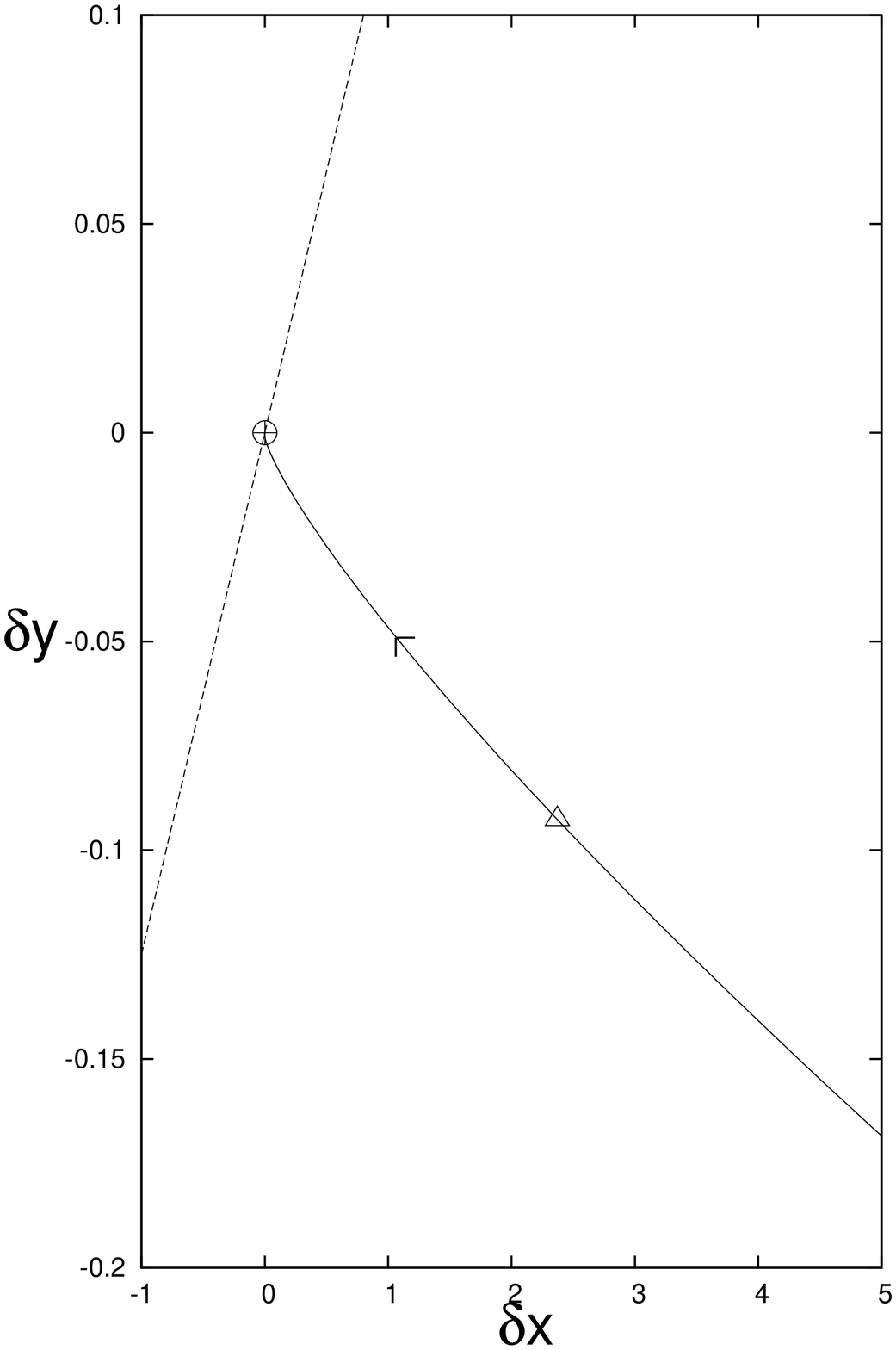}
\includegraphics[width=5.0cm]{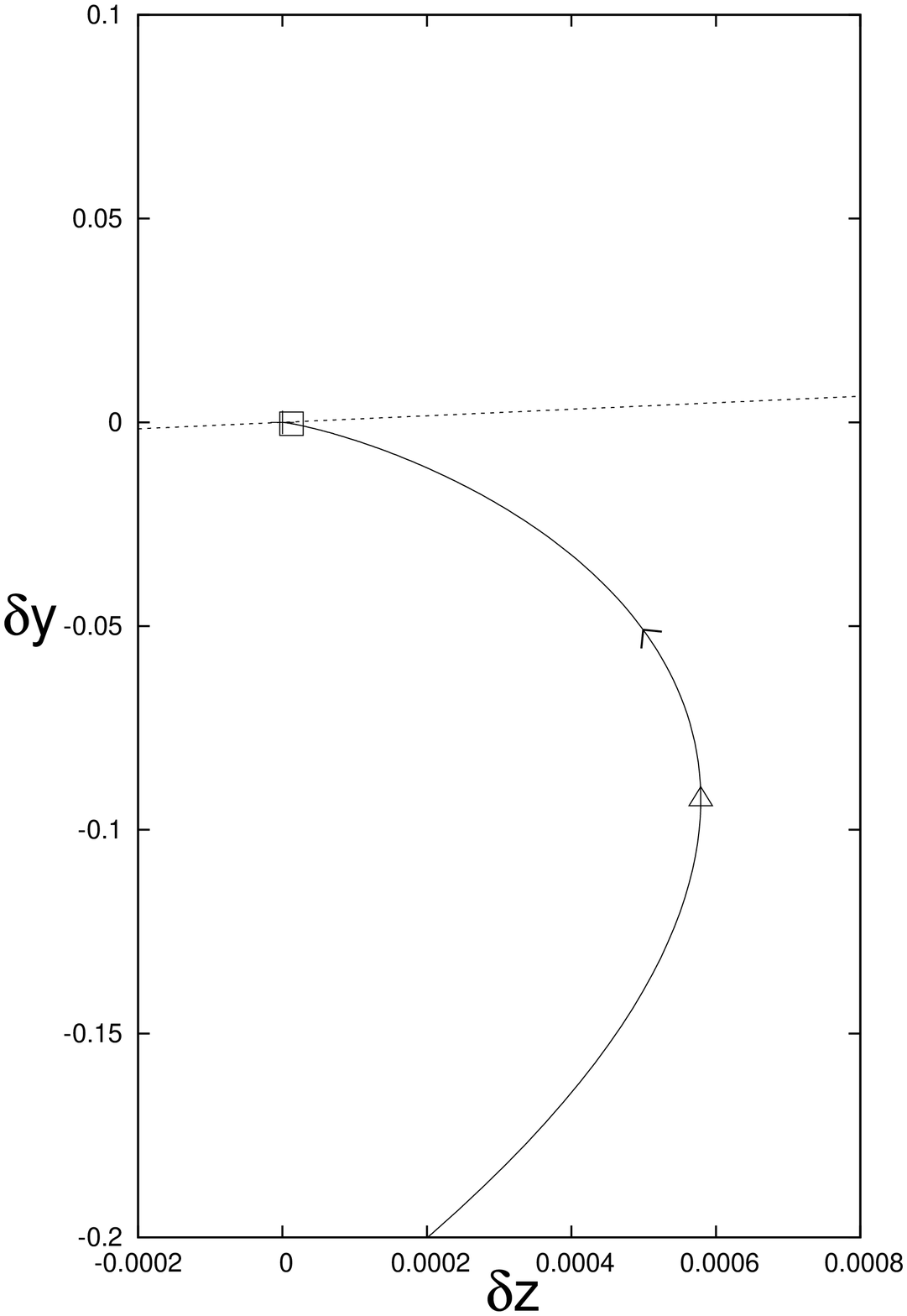}
\includegraphics[width=5.0cm]{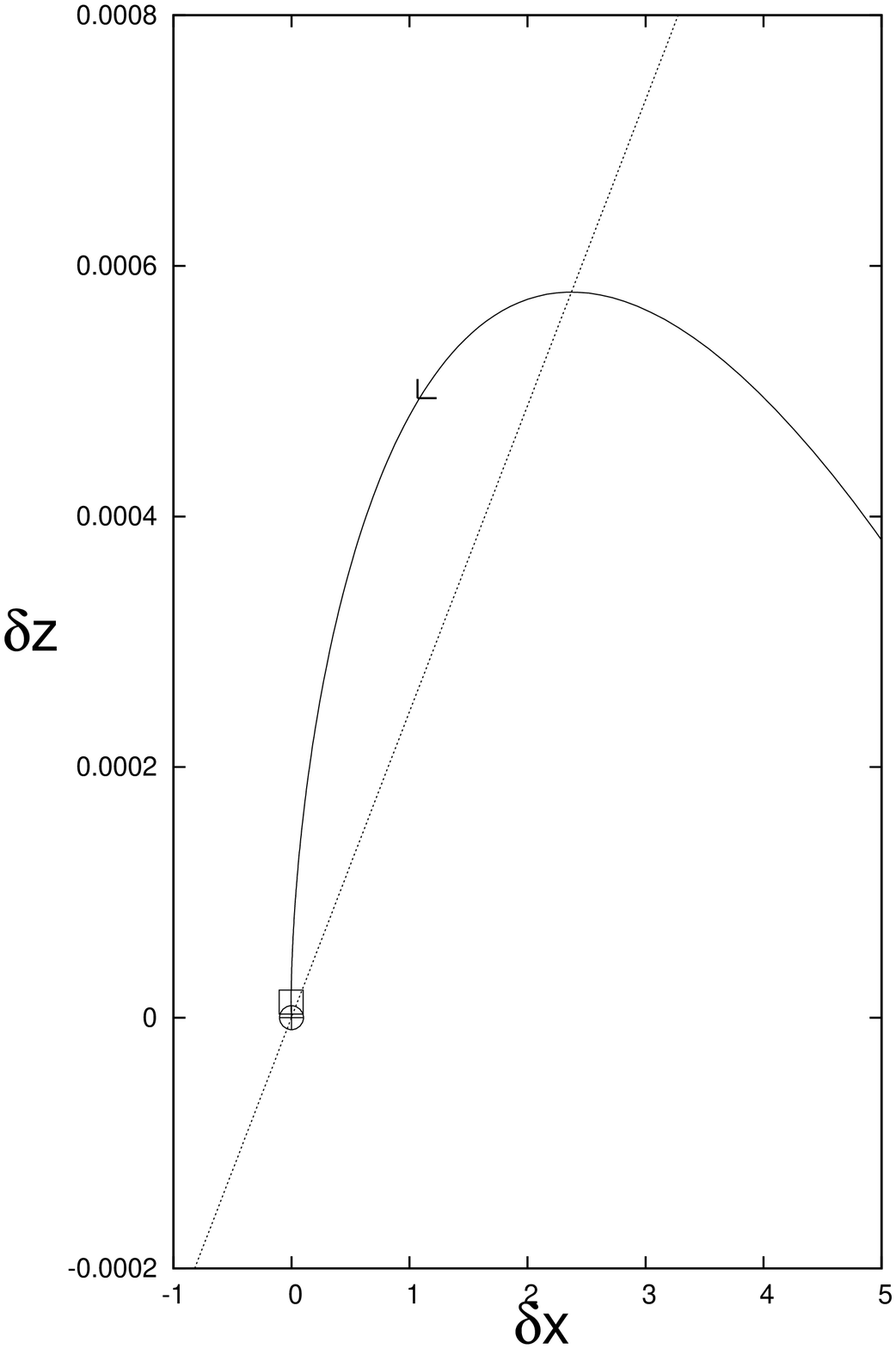}
\includegraphics[width=5.0cm]{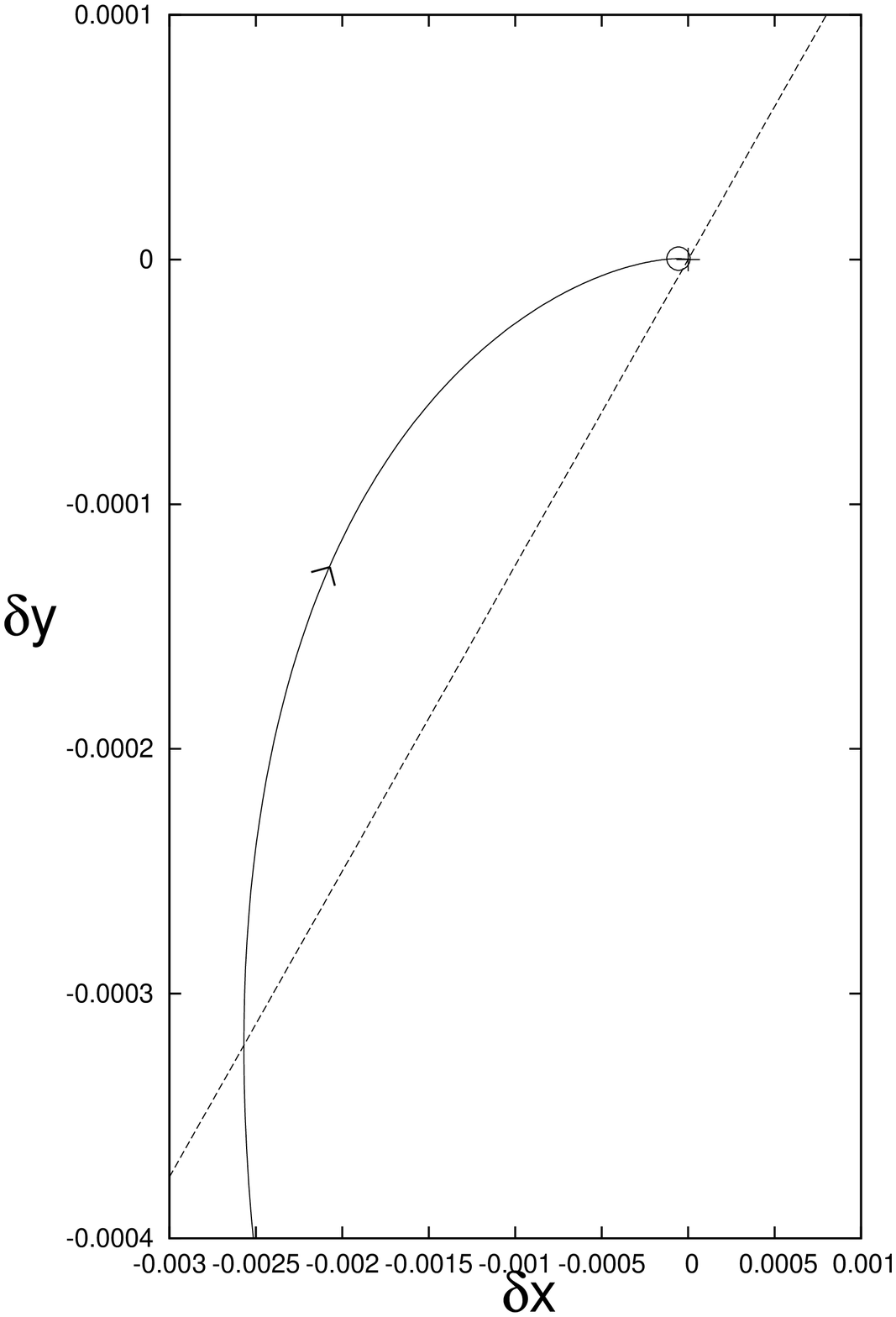}
\includegraphics[width=5.0cm]{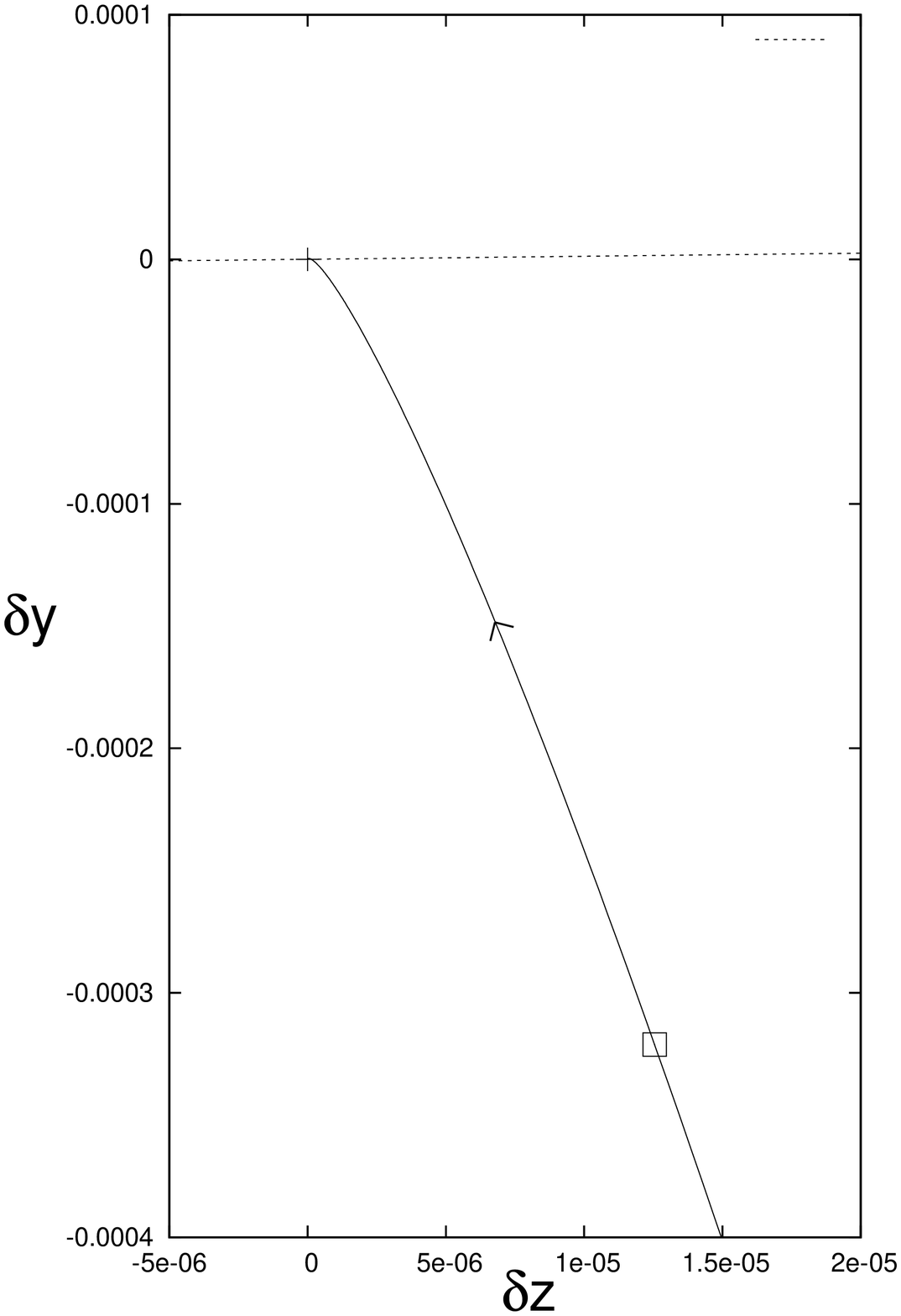}
\includegraphics[width=5.0cm]{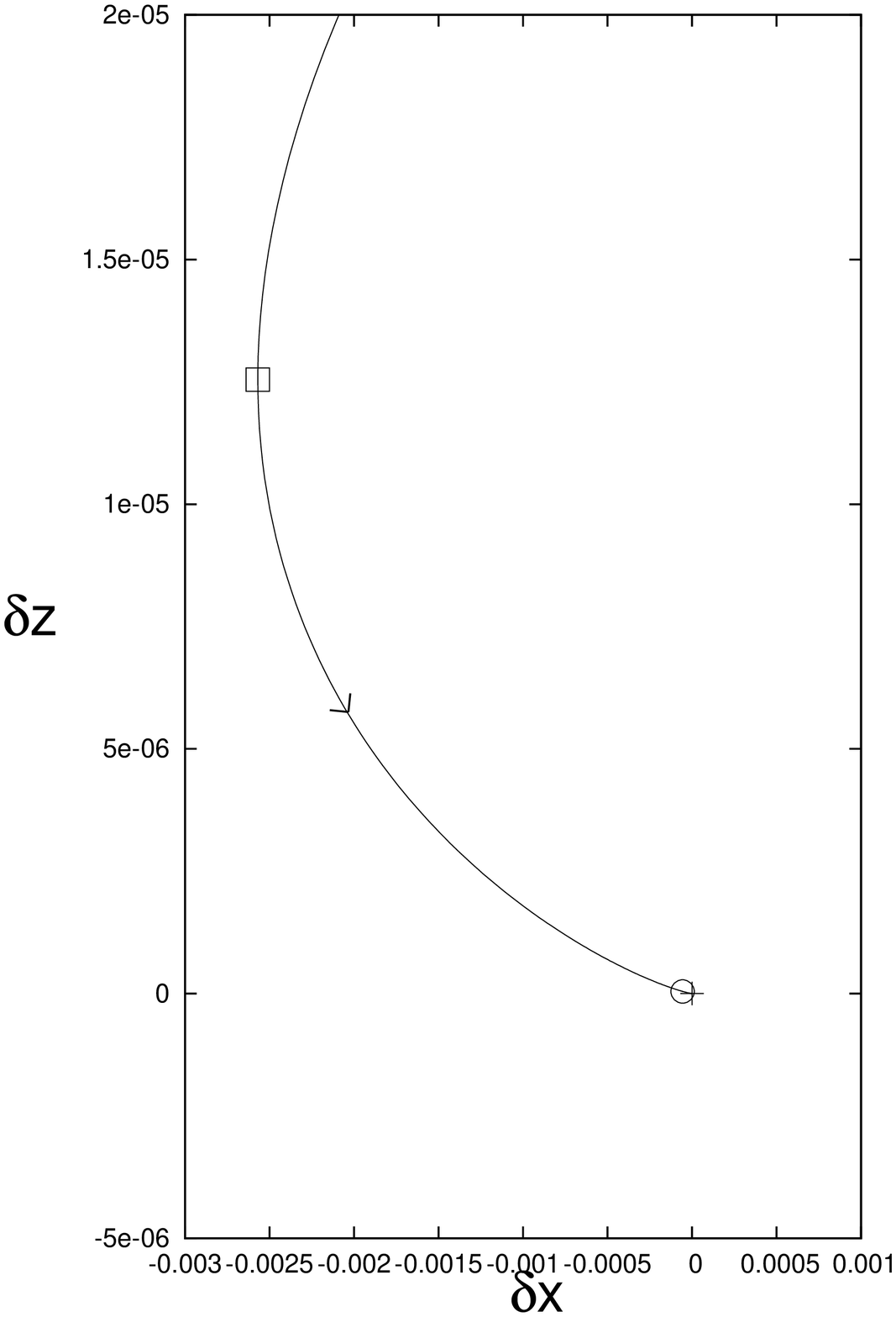}
\includegraphics[width=5.0cm]{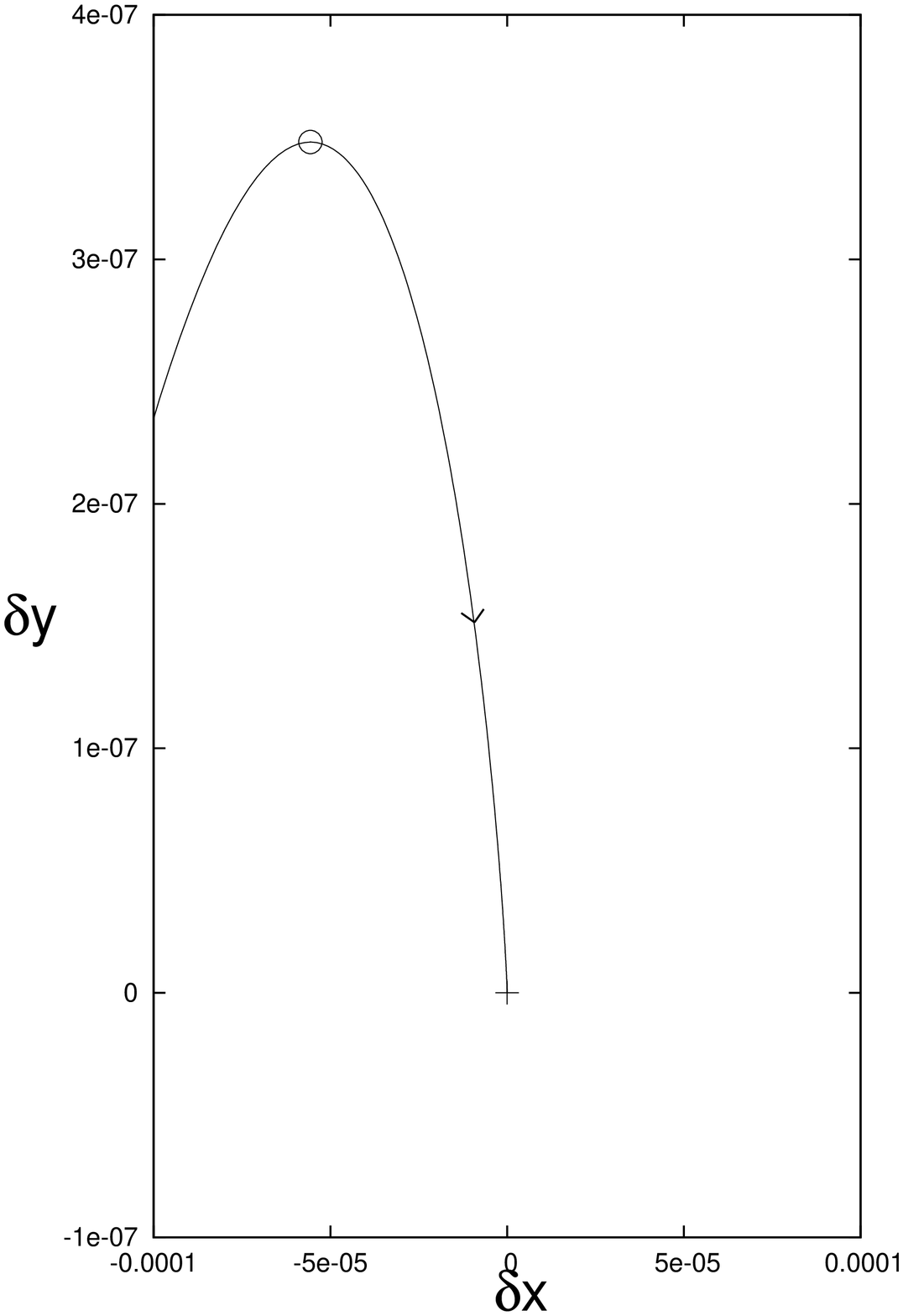}
\includegraphics[width=5.0cm]{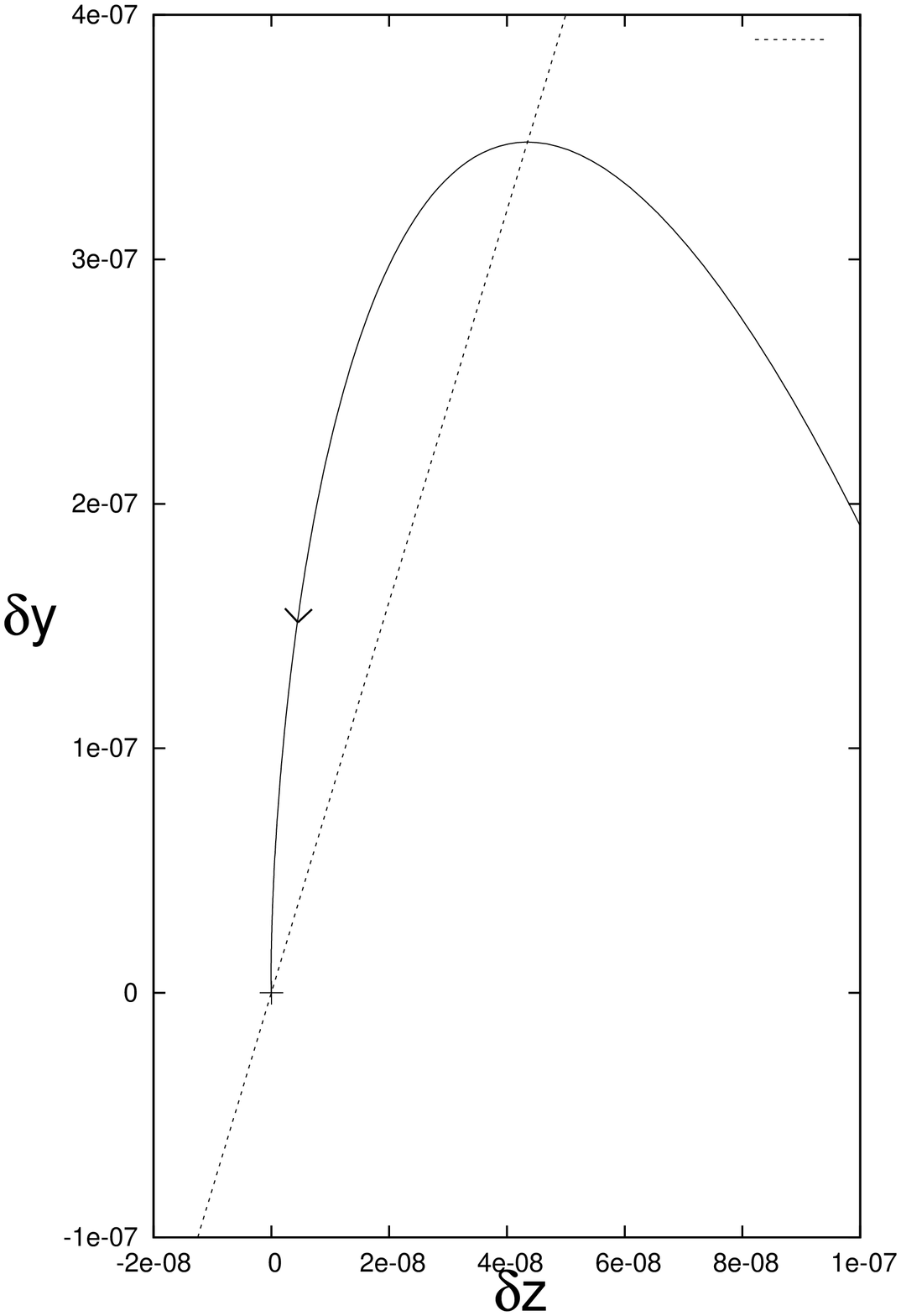}
\includegraphics[width=5.0cm]{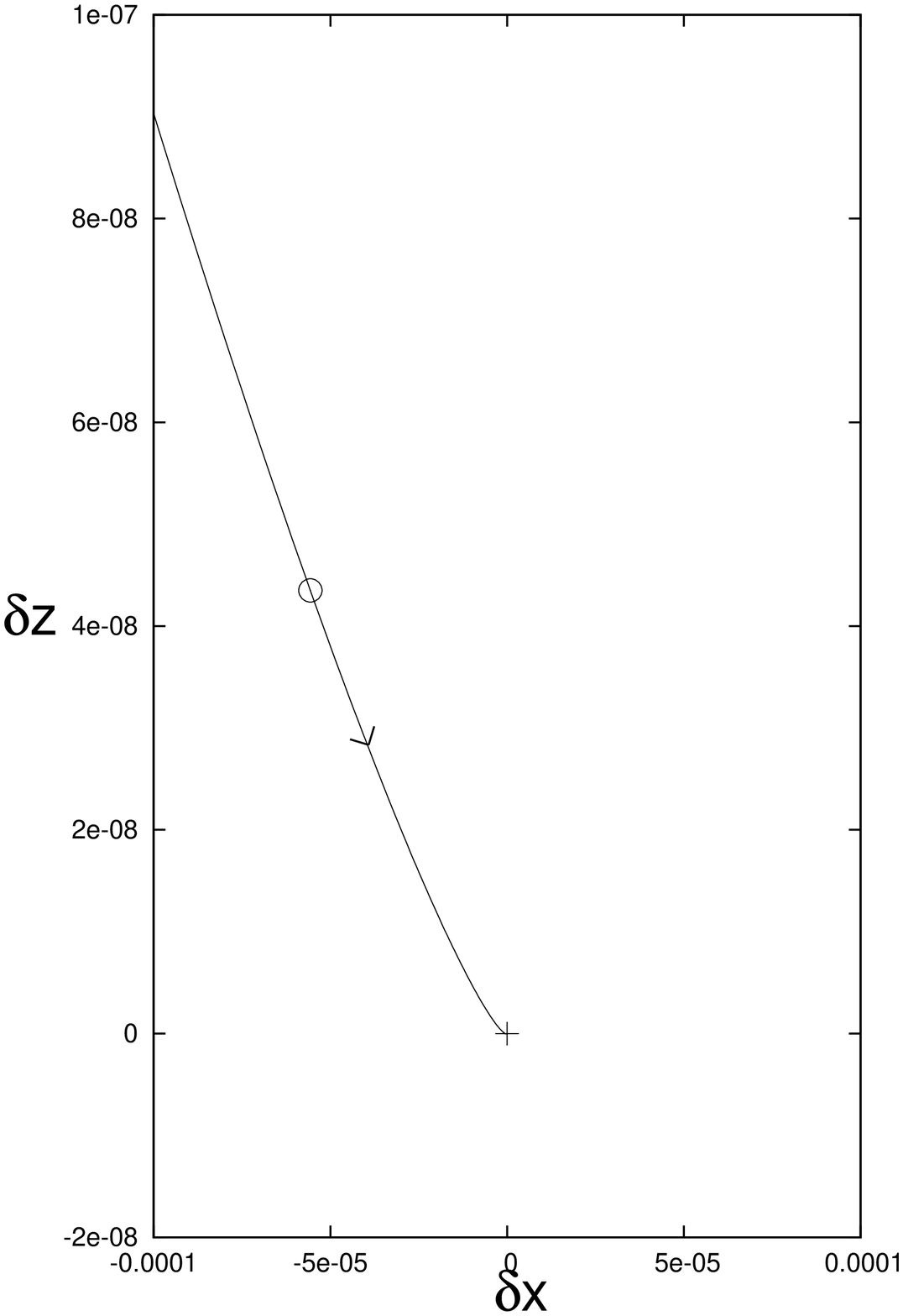}
\caption{Projection of the asymptotic trajectory to (a) $x$-$y$ plane (front view), (b) $z$-$y$ plane (left view), and (c) $x$-$z$ plane (bottom view). The projections of nullclines are shown in dashed lines whenever possible. See the text for more details.
%Otherwise, only the intersection points with the trajectory are indicated, by open square ($x$-nullcline), open circle ($y$-nullcline), and open triangle ($z$-nullcline). Figures (d),(e),and (f) are the magnification of (a), (b), and (c) near the fixed point. Figures (g), (h), and (i) are still larger magnifications of (a), (b) and (c).
}\label{traj}
\end{figure}

%Since we have four parameters in $\alpha$, $\beta$, $\gamma$, and $\nu$, the region with $\Delta <0$ has to be drawn in a four-dimensional space and thus not easy to visualize directly. Also, although the nonzero fixed point $\tilde x$ is a well-defined function of these four parameters, it has no analytic formula for $m>2$, making the task more difficult. However, one can always perform a numerical scan to check to find the values of the parameters that belong to this region. 
%Here, let us consider two-dimensional sections in the $\alpha-\beta$ space, for particular values of $m$, $\gamma$ and $\nu$. Let us consider the simplest case $m=$. We first put $\nu=0$, which is in fact negligible for most practical applications as was explained earlier. When $\gamma=0$, we see that $\Delta=(\alpha-beta)^2 \alpha^2 \beta^2 > 0$ the region of $\Delta <0$ vanishes. We have also seen earlier that $\Delta>0$ also for large enough $\gamma$.  
 As an example, let us consider $m=1$ with $\alpha=\beta=1/8$, $\gamma=1/2^9$, $\nu=0$.  We have a nonzero stable fixed point at $x_*=1/\alpha\beta-1=63$, $y_*=x_*/(\alpha(1+x_*)) = 63/8$, $z_*=x_*/(1+x_*) = 63/64$. Since $\bn=0$, $\bb=1/8$, $\bar \gamma = 1/8$, $\zeta=1/2^{15}$, we have
\be
\Delta  &=& (\alpha - \bb)^2 (\bb-\bg)^2 (\bg-\alpha)^2 - 2 (\alpha + \bb - 2 \bg)(\bb+\bg-2\alpha)(\bg+\alpha-2\bb)\tg\nonumber\\
&& -27 \tg^2 + 18 (\bb+\bg- 2 \alpha) \tg^2 \bn\nonumber\\
&& + 2 (\alpha - \bb)(\alpha - \bg) (2 \alpha^2 - \bb^2 - \bg^2 - 2 \alpha \bb + 4 \bb \bg - 2 \bg \alpha) \tg \bn\nonumber\\
&& + (-8 \alpha^2 + \bb^2 + \bg^2 + 8 \alpha \bb + 8 \bg \alpha - 10 \bb \bg ) \tg^2 \bn^2 + 4 \tg^3 \bn^3\nonumber\\
&=&  - 27(\frac{1}{2^{15}})^2 < 0,
\ee
leading to one real root and two complex roots for the characteristic equation Eq.(\ref{3ch}):
\be
\lambda_0 &=& -\frac{3}{32}\nonumber\\
\lambda_\pm &=& \frac{-9 \pm i \sqrt{3}}{64}.
\ee
with the corresponding eigenvectors
\be
{\bf u}_0^T = (32,1,1/32),\quad {\bf u}_\pm^T = (64,-1 \pm i \sqrt{3},(-1 \mp i \sqrt{3})/32),\quad 
\ee
leading to
\be
{\bf u_1}^T = (128,-2,-1/16),\quad {\bf u_2}^T = (0,2\sqrt{3},-\sqrt{3}/16) 
\ee
and the asymptotic behavior
\be
\left(\begin{array}{c} x(t) \\ y(t) \\ z(t)\end{array}\right) = \left(\begin{array}{c} 63 +32 A e^{-3t/32} + 128 B e^{-9 t/64}  \cos (\frac{\sqrt{3} }{64}(t + \phi))  \\ \frac{63}{8} +  A e^{-3t/32} - 2 B e^{-9 t/64} \left( \cos (\frac{\sqrt{3} }{64}(t + \phi)) +  \sqrt{3} \sin (\frac{\sqrt{3} }{64}(t + \phi)\right) \\ \frac{63}{64} + \frac{1}{32} A e^{-3t/32} - \frac{1}{16}  B e^{-9 t/64}\left( \cos (\frac{\sqrt{3} }{64}(t + \phi)) -  \sqrt{3}  \sin (\frac{\sqrt{3} }{64}(t + \phi))\right) \end{array}\right)\label{asym}
\ee
This behavior can be understood using nullclines, just as in the case of two-dimensional system\cite{Stro}. Note that in the case of the three-dimensional system, a nullcline is a two-dimensional surface. Assuming $\nu=0$ for simplicity, we see from eq.(\ref{3d})  that the nullclines for $x$,$y$, and $z$ are the surfaces $y= \beta x$, $z = \alpha y$, and $z = \frac{x^m}{x^m + 1}$, respectively. Linearizing the equation around the fixed points, we get the planes:
\be
\delta y &=& \beta \delta x\nonumber\\
\delta z &=& \alpha \delta y\nonumber\\
\delta z &=& \frac{m x_*^{m-1}}{(x_*^m + 1)^2} \delta x.
\ee
Part of the trajectory Eq.(\ref{asym}) with $A=\phi=0$ and $B=1$ is shown in Fig.\ref{traj}.
The projection to $x$-$y$, $z$-$y$, and $x$-$z$ planes are shown in Figure \ref{traj} (a),(b), and (c), which may be considered as front, left, and bottom view. The fixed point is denoted by a plus symbol at the origin. The projection of $x$-nullcline, $y$-nullcline, and the $z$-nullcline are indicated in Figures(a),(b), and (c) as dashed lines. The intersections of the trajectory with $x$-nullcline, $y$-nullcline, and the $z$-nullcline are indicated with the open square, open circle, and open triangle respectively, in the figures where the projection of the corresponding nullclines cannot be drawn as lines. 

Initially the concentrations of the protein X is  decreasing, that of RNA is increasing, and the probability of the gene being in the active state is increasing. As can be shown in Figure \ref{traj} (c), eventually the trajectory crosses  the $z$-nullcline, indicating that the protein level has decreased to a level where probability of the gene being in the active state begins to decrease. The intersection point is denoted by triangles in Figure \ref{traj} (a) and (b). The crossing of the $x$ and $y$-nullcline is invisible with this resolution, and the protein level $x$ keeps decreasing and the RNA level $y$ keeps increasing  until the trajectory goes very close to the fixed point. Figures (d),(e), and (f) are the magnification of (a),(b) and (c) near the fixed point. As can be seen from Fig.\ref{traj} (d), the trajectory indeed crosses the $x$-nullcline, indicating that the RNA has been increased to a level so that it can produce more proteins than being degraded, so the protein level $x$ begins to increase. The intersection is denoted as open squares in figures (e) and (f). The crossing of $y$-nullcline is invisible in this resolution, and the concentrations of both protein and RNA increase, and the probability of the gene being in the active state decreases, until the system goes closer to the fixed point. Figures (g), (h), and (i) are the magnification of (d),(e) and (f) near the fixed point. As can be seen from Fig.\ref{traj} (h), the trajectory indeed crosses the $y$-nullcline, indicating that the probability of gene being in the active state has decreased so much that the production of RNA is less than that being degraded, and consequently the RNA concentration begins to decrease. Similar cyclic behavior will repeat in smaller scales as the trajectory converges to the fixed point.

Similar oscillatory behavior exists near a saddle point when the Hessian has complex roots, although the concentrations will eventually diverge away from the fixed point and the linear approximation will break down. 
\section{Conclusion}
In this work, I analyzed the stability of the fixed points of the simplest auto-regulatory system, consisting of DNA, RNA, and the protein product that positively regulate its production upon binding to the gene. After linearizing the equation around fixed points, the eigenvalues of the three-dimensional Hessian were obtained using the root formula of the cubic equation. The result rigorously confirms the well-known results from the two-dimensional analysis: When Hill coefficient is one there is only one stable fixed point, that with zero concentrations when the decay rates of the RNA and the protein product are large compared to their production rate, and that with nonvanishing concentration if the production rates are large compared to the decay rates, zero concentrations being an unstable fixed point in this case. On the other hand, for Hill coefficient larger than one, the vanishing concentration is always a stable fixed point, and an additional nonzero stable fixed point appears when the production rates are large compared to the decay rates.

Although the general conclusion about the stability of the fixed points are the same as the two-dimensional analysis, we could obtain more detailed information such as the number of positive eigenvalues near a saddle point, and the number of complex eigenvalues near an arbitrary fixed point. In particular, we could see that the approach to the fixed point in the three-dimensional system can exhibit oscillatory behavior for certain values of the parameters, a property that cannot be observed in the two-dimensional system. In fact, the two-dimensional system can be justified only in the limit of fast equilibration of gene, in which case the eigenvalues are all real. On the other hand, the novel oscillatory behavior appears when binding and unbinding of protein to the DNA is slow.

Although the analysis is much more general than the two-dimensional analysis, the deterministic equation (\ref{deter}) is valid only when the concentration of protein is large enough so that its relative fluctuation can be neglected. Eventually the fluctuations will grow with time and will no more be negligible. Also, the stochastic effect cannot be neglected near the origin where the number of proteins and RNA are expected to be small. Therefore, the full master equation or Fokker-Planck equation will have to be considered for more accurate analysis\cite{Haoge1,Haoge2,Haoge3,Lip,Wangjin1,Wangjin2}. Moreover, the current formalism neglects the time delay between the transcription and translation, and that between the translation and the positive regulation. The current formalism is expected to work well for the regulation process in prokaryotes where all the processes happen in more or less the same region in space, and will be less accurate for Eukaryote where the process of transcription and translation  happen at distant locations. The effect of time delay due to finite diffusion coefficients will have to be considered for a more accurate description of such systems\cite{Wangjin1,Wangjin2}.
\section{Acknowledgements}
This work was supported by the
National Research Foundation of Korea, funded by the Ministry of Education, Science, and Technology (NRF-2012M3A9D1054705).
%\newpage
\appendix
\section{Deviation of the rate equations}
Let us denote the numbers of proteins, RNA molecules, and active genes as $N_X$,$N_Y$, and $N_Z$ respectively. We assume there is only one copy of the gene, so $N_Z=0$ or $N_Z=1$. Denoting the probability for $N_X=x$, $N_Y=y$, and $N_Z=z$ as $P(x,y,z)$, the master equations corresponding to the process Eq.(\ref{rateeq}) is
%\be
%\dot P(X,Y,1) &=& \tilde k_1 P(X+m,Y,0)(X+m)^m - k_2 P(X,Y,1) + a P(X,Y-1,1) - a P(X,Y,Z)\nonumber\\
%&&+ c P(X-1,Y,1) Y - c P(X,Y,1) Y + b P(X,Y+1,1) (Y+1) - b P(X,Y,1) Y\nonumber\\
%&& + d P(X+1,Y,1)(X+1) - d P(X,Y,1) X \nonumber\\
%\dot P(X,Y,0) &=& - k_1 P(X,Y,0) X^m + k_2 P(X-m,Y,1)  + c P(X-1,Y,0) Y - c P(X,Y,0) Y\nonumber\\
%&&+ b P(X,Y+1,0) (Y+1) - b P(X,Y,0) Y\nonumber\\
%&&+ d P(X+1,Y,0)(X+1) - d P(X,Y,0) X. \label{master}
%\ee
%where $P(X,Y,Z)$ is the probability that the number of protein is $X$,  RNA $Y$, and that the number of active gene is $Z\quad (=0,1)$. Eq.(\ref{master}) can be compactly rewritten as
\be
\dot P(x,y,z) &=& \tilde k_1 P(x+m,y,1-z)(x+m)^m z  - \tilde k_1 P(x,y,z) x^m (1-z) \nonumber\\
&&-  k_2 P(x,y,z) z  +   k_2 P(x-m,y,1-z) (1-z) +   a P(x,y-1,z)z - \ a P(x,y,z) z\nonumber\\
&&+  c P(x-1,y,z) y -  c P(x,y,z) y +  b P(x,y+1,z) (y+1) -  b P(x,y,z) y\nonumber\\
&& +  d P(x+1,y,z)(x+1) -  d P(x,y,z) x \label{master2}
\ee
where $\tilde k_1 \equiv  k_1/(c_0 V)^m$ is used as the rate constant for binding of the protein $X$ to the gene,  where V is the volume of the region where the process is taking place and $c_0$ is a standard reference concentration. Usage of $\tilde k_1$ here ensures that we get binding rate constant $k_1$ when the equations are expressed in terms of $[X]$, $[Y]$, and $[Z]$.
The equations describing the time evolution of the expectation values of $N_X$, $N_Y$, and $N_Z$ can be obtained by multiplying Eq.(\ref{master2}) by $x$, $y$, and $z$ and taking summation:
\be
\langle \dot N_Z \rangle = \sum_{x,y,z} z \dot P(x,y,z) &=& \sum_{x,y,z} [\tilde k_1 P(x+m,y,1-z)(x+m)^m z^2  - \tilde k_1 P(x,y,z) x^m z (1-z) \nonumber\\
&&- k_2 P(x,y,z) z^2  +  k_2 P(x-m,y,1-z) (1-z)z ] \nonumber\\
%&=& \sum_{x,y,z}\left[k_1 P(x+m,y,1-z)(x+m)^m z  - k_2 P(x,y,z) z  \right] \nonumber\\
&=&  \tilde k_1 \sum_{x,y,z} P(x,y,z)x^m (1-z)  - k_2 \sum_{x,y,z} P(x,y,z) z  \nonumber\\
&&-\tilde k_1 \sum_{x < m} P(x,y,z) x^m (1-z)^2   \nonumber\\
&=&\tilde k_1 \langle N_X^m (1-N_Z) \rangle - k_2 \langle N_Z \rangle \nonumber\\
&&-\tilde k_1 \sum_{x < m} P(x,y,z) x^m (1-z)   \label{pzeq}
\ee
\be
\langle \dot N_Y \rangle = \sum_{x,y,z} y \dot P(x,y,z) &=& a \sum_{x,y,z} [ P(x,y,z) (y+1) z -  P(x,y,z) y z] \nonumber\\
&& + b \sum_{x,y,z} [ P(x,y,z) y(y-1) - P(x,y,z) y^2 ] \nonumber\\
&& -\tilde k_1 \sum_{x < m} P(x,y,1-z) x^m y z^2 \nonumber\\
&=& a\sum_{x,y,z} P(x,y,z) z - b \sum_{x,y,z} P(x,y,z) y \nonumber\\
&& -\tilde k_1 \sum_{x < m} P(x,y,z) x^m y (1-z)^2 \nonumber\\
&=& a \langle N_Z \rangle - b \langle N_Y \rangle \nonumber\\
&& -\tilde k_1 \sum_{x < m} P(x,y,z) x^m y (1-z) \label{pyeq}
\ee
\be
\langle \dot N_X \rangle = \sum_{x,y,z} x \dot P(x,y,z) &=& \sum_{x,y,z}  [\tilde k_1 P(x+m,y,1-z)(x+m)^m x z - \tilde k_1 P(x,y,z) x^{m+1} (1-z) \nonumber\\
&&- k_2 P(x,y,z) x z  +  k_2 P(x-m,y,1-z) x (1-z)  \nonumber\\
&&+ c P(x-1,y,z) x y - c P(x,y,z) x y\nonumber\\
&&+ d P(x+1,y,z)(x+1)x - d P(x,y,z) x^2 ]\nonumber\\
&=& - m \tilde k_1 \sum_{x,y,z} P(x,y,z) x^m z + m k_2 \sum_{x,y,z} P(x,y,z) z \nonumber\\
&&+c \sum_{x,y,z} P(x,y,z) y - d \sum_{x,y,z} P(x,y,z) x \nonumber\\
&& - \sum_{x < m}  \tilde k_1 P(x,y,1-z) x^m (x-m) z \nonumber\\
&=& - m \tilde k_1 \langle N_X^m N_Z \rangle + m k_2 \langle N_Z \rangle + c \langle N_Y \rangle -d \langle N_X \rangle \nonumber\\
&& - \sum_{x < m}  \tilde k_1 P(x,y,z) x^m (x-m) (1-z) \label{pxeq}
\ee
where we used the fact $z^2=z$ and $z(1-z)=0$ since $z=0$ or $z=1$. 
%Summing over $X$ and $Y$, we get the equation for the marginal probability for the gene states:
%\be
%\dot P(Z=1)&=&k_1 \sum_X X^m P(X,Z=1) - k_2 P(Z=1) \nonumber\\
%\dot P(Z=0)&=&-k_1 \sum_X X^m P(X,Z=0) + k_2 P(Z=1).
%\ee
%Note that, since $Z=0$ or $Z=1$, $P(Z)=<Z>$ and $\sum_X f(X) P(X,Z) = <f(X)Z$ where an angular bracket denotes the expectation value. Also the sum of these two equations are zero, so they contain the same information. Therefore the equations above can be rewritten as
%\be
%\dot <Z>&=&k_1 <X^m (1-Z)> - k_2 <Z>.
%\ee 
When $\langle N_X \rangle$ is so large that its fluctuation is small relative to the average value $\langle N_X \rangle$, we may make approximations
\be
\langle N_X^m N_Z \rangle &\simeq& \langle N_X \rangle^m \langle N_Z \rangle,\nonumber\\
\langle N_X^m (1-N_Z) \rangle &=& \langle N_X^m \rangle -\langle N_X^m N_Z \rangle \simeq \langle N_X^m \rangle - \langle N_X \rangle^m \langle N_Z \rangle = \langle N_X \rangle^m \langle 1-N_Z \rangle.
\ee
After dividing by $c_0 V$,  last terms in Eqs.(\ref{pzeq}), (\ref{pyeq}), and (\ref{pxeq}) becomes negligible in the limit of $V \to \infty$. Defining $[A]\equiv \langle N_A \rangle /(c_0 V)$,  the rate equation (\ref{deter}) is now derived.  
\section{Fixed points}
In this section, I first review how the fixed points of the system (\ref{2d}), or equivalently those of Eq.(\ref{3d}), are determined. In order to get the fixed points, we first solve the equation for the nullcline $\dot y = 0$, to get
\be
y=\frac{x^m}{\alpha(1+x^m)}
\ee 
and substitute into the nullcline equation for $x$, $\dot x = 0$, which after some manipulation becomes
\be
x \left( \alpha \beta x^m -  x^{m-1} + \alpha \beta \right) = 0
\ee
We see that $x=0$ is always a fixed point, and there can be additional fixed points which are the roots of the equation
\be
f(x) \equiv x^m - \frac{1}{\alpha \beta} x^{m-1} + 1 = 0 \label{theeq}
\ee
for certain values of $\alpha$ and $\beta$.

When $m=1$, there is one nonzero real root to the equation (\ref{theeq}) if and only if
\be
\alpha \beta < 1 \label{fixeq1}
\ee
For  $m>1$, there can be up to two real roots. In order to investigate the existence of such roots, we first locate the extremum of $f(x)$, by taking its derivative and setting it to zero:
\be
x^{m-2} \left(m x - \frac{ m-1}{\alpha \beta}\right) = 0
\ee
yielding one extremum at 
\be
x_1 = \frac{m-1}{ m \alpha \beta}
\ee
and the other one at $x=0$ if $m > 2$. Since $f^{(j)}(0) = 0$ for $j<m-1$ and $f^{(m-1)}(0)=-(m-1)!/\alpha\beta < 0$, where $f^{(n)}$ denotes the $n$-th derivative of $f(x)$, we see that $x=0$ for $m>2$ is a local maximum. On the other hand, since
\be
f^{(2)}(x_1) &=& m(m-1)\left(\frac{m-1}{m  \alpha \beta}\right)^{m-2}\nonumber\\
&&- \frac{(m-1)(m-2)}{\alpha\beta} \left(\frac{m-1}{ m \alpha \beta}\right)^{m-3}\nonumber\\
&=&m \left(\frac{m-1}{m \alpha \beta}\right)^{m-2} > 0, 
\ee
 we see that $x_1$ is the global minimum of $f(x)$. Since $f(0)=1 > 0$ and $\lim_{x \to \infty} f(x) = \infty > 0$, Eq. (\ref{theeq}) will have two distinct real roots if and only if the value of $f(x)$ negative at the global minimum $x_1$:
\be
f(x_1) &=&  \left(\frac{m-1}{m \ab}\right)^m - \frac{1}{\ab}\left(\frac{m-1}{m \ab}\right)^{m - 1} + 1 \nonumber\\
&=& -\left(\frac{1}{m-1}\right)\left(\frac{m-1}{m \ab}\right)^{m} + 1\nonumber\\
&& < 0
\ee
which can be rewritten as
\be
\left( \frac{m-1}{m \alpha \beta}\right)^m > m-1. \label{fixeq2}
\ee

\section{The stability analysis of the two-dimensional system}
Let us consider the stability of the two-dimensional system Eq.(\ref{2d}) around the fixed points. Most of the material can be found in previous literature\cite{Stro,Griff,Gri68}, but I write down an explicit form of eigenvalues so that they can be compared with those obtained from the three-dimensional system in the limit of ultra-fast binding and unbinding of the protein to the gene. Considering a small deviation of $x$ and $y$ around a fixed point $(x_*,y_*)$ of Eq (\ref{2d}), with $x=x_* + \delta x$,$y=y_* + \delta y$, we get a linearized equation
\be
\left(\begin{array}{c} \delta x\\ \delta y\end{array}\right) = {\bf A} \left(\begin{array}{c} \delta x\\ \delta y\end{array}\right) = \begin{pmatrix} -\beta  & 1\\ \frac{m x_*^{m-1} }{(1+x_*^m)^2} & -\alpha \end{pmatrix} \left(\begin{array}{c} \delta x\\ \delta y\end{array}\right)
\ee 
Solving the characteristic equation for the matrix $A$,
\be
|\lambda - {\bf A}| &=& (\lambda + \alpha) (\lambda + \beta) - \frac{m x_*^{m-1}}{(1+x_*^m)^2}  \nonumber\\
&=& \lambda^2 + (\alpha + \beta ) \lambda + \alpha \beta -  \frac{m x_*^{m-1}}{(1+x_*^m)^2} = 0,
\ee 
we get the eigenvalue
\be
\lambda_{\pm} &=& \frac{1}{2}\left[ - \alpha - \beta \pm \sqrt{\left(\alpha + \beta \right)^2- 4 \alpha \beta + 4 \frac{m  x_*^{m-1}}{(1+x_*^m)^2} } \right]\nonumber\\
&=& \frac{1}{2}\left[ - \alpha - \beta \pm \sqrt{\left(\alpha - \beta \right)^2+ 4 \frac{m  x_*^{m-1}}{(1+x_*^m)^2} } \right]. \label{ev2}
\ee
Since the expression inside the square root is nonnegative, we see that the eigenvalues are real. 

From the first line of Eq.(\ref{ev2}), we see that if
\be
\frac{m  x_*^{m-1}}{(1+x_*^m)^2}  < \ab, \label{stab}
\ee
the absolute value of the square root term is smaller than the first term which is negative, so signs of both eigenvalues are negative and we get a stable fixed point. On the other hand, if
\be
\frac{m  x_*^{m-1}}{(1+x_*^m)^2}  > \ab, \label{sadd}
\ee
the sign of the eigenvalues are determined by the square root term, so both positive and negative signs appear, and we get a saddle point. It is straightforward see that the equality corresponds to the marginal case where one of the eigenvalues becomes zero. This general result can be used for analyzing the stability of the specific fixed points.

\subsection{$m=1$}
\subsubsection{$\ab > 1$}
 As discussed in the appendix B, $x_*=0$ is a unique fixed point in this case. The condition $\ab > 1$ is the same as (\ref{stab}) with $x_*=0$, so this is a stable fixed point.
\subsubsection{$\ab < 1$}
It is clear that $x_*=0$ is now a saddle point, since now (\ref{sadd}) holds with $x_*=0$. There is an additional fixed point which is a solution to the equation (\ref{theeq}):
\be
x_*=\frac{1}{\ab} -1.
\ee
We see that
\be
\frac{m x_*^{m-1}}{(1+x_*^m)^2} = \frac{1}{(1+x_*)^2}= (\ab)^2 < \ab.
\ee
Therefore, the condition (\ref{stab}) is satisfied and we get a stable fixed point.

To summarize, for $m=1$, when the RNA and protein is degraded fast enough so that $\ab > 1$, $x_*=0$ is the unique stable fixed point of (\ref{2d}). Eventually the concentrations of both species go to zero. On the other hand, when the degradation is slow enough so that $\ab < 1$, then $x_*=0$ is a saddle point and an additional stable nonzero fixed point $x_*=\frac{1}{\ab} -1$ appears. We see that there is no bistability for $m=1$.
\subsection{$m > 1$}
In this case, $x_*=0$ is always a stable fixed point since 
\be
\frac{m x_*^{m-1}}{(1+x_*^m)^2} = 0 < \ab.
\ee
As discussed in the appendix B, when Eq.(\ref{fixeq2}) holds, we have additional nonzero fixed points which are the two distinct real roots $x_- < x_+$ of the equation (\ref{theeq}). After multiplying $\alpha \beta (1 + x_*^m)^2$ to both sides of  Eq.(\ref{stab}) and Eq.(\ref{sadd}), we see that we get stable fixed point or saddle point depending on whether the sign of  the expression
\be
\ab m x_\pm^{m-1} - (\ab)^2 (1 + x_\pm^m)^2 = x_\pm^{m-1}(m \ab  -  x_\pm^{m-1})
\ee
is negative or not, where Eq.(\ref{theeq}) was used in going from the second to the third expression. Therefore, we only have to examine the sign of 
\be
m \ab - x_\pm^{m-1}.
\ee
Let us define 
\be
g(x) \equiv m \ab - x^{m-1}.
\ee  
we see that this function is zero when
\be
x=x_0 \equiv (m \ab)^{1/(m-1)}.
\ee
On the other hand, we have
\be
f(x_0) = \left( m\ab\right)^{m/(m-1)} - (m-1) < 0
\ee
where the last inequality follows from the condition (\ref{fixeq2}). Therefore, we see that 
\be
x_- < x_0 < x_+.
\ee
Since $g(x)$ is a monotonically decreasing function, it immediately follows that $g(x_-) > 0$ and $g(x_+) < 0$. Therefore, $x_-$ is a saddle point and $x_+$ is a stable fixed point. 
To summarize, when RNA and the protein are degraded fast enough so that 
\be
\left( \frac{m-1}{m \alpha \beta}\right)^m < m-1,
\ee
$x=0$ is a unique sable fixed point. When the degradation is slow enough so that the inequality is inverted, then we have additional nonzero fixed points $x_- < x_+$, where $x_-$ is a saddle point and $x_+$ is a sable fixed point. Since $x=0$ and $x=x_+$ are both stable fixed points, we get bistability.

\end{document}